# Evaluation of Measurement Comparisons Using Generalised Least Squares: the Role of Participants' Estimates of Systematic Error

John F. Clare[a*], Annette Koo[a], Robert B. Davies[b]

[a]Measurement Standards Laboratory of New Zealand, Industrial Research Ltd, P.O. Box 31-310, Lower Hutt, New Zealand
[b]Statistics Research Associates Ltd, P.O. Box 12-649, Wellington, New Zealand

ABSTRACT

We consider the evaluation of laboratory practice through the comparison of measurements made by participating metrology laboratories when the measurement procedures are considered to have both fixed effects (the residual error due to unrecognised sources of error) and random effects (drawn from a distribution of known variance after correction for all known systematic errors). We show that, when estimating the participant fixed effects, the random effects described can be ignored. We also derive the adjustment to the variance estimates of the participant fixed effects due to these random effects.

* Corresponding Author.
  E-mail addresses: j.clare@irl.cri.nz, a.koo@irl.cri.nz, robert@statsresearch.co.nz

# 1. Introduction

National Metrology Institutes are responsible for independently realising or establishing scales of physical measurement within a nation. These laboratories participate in 'key comparisons' to compare their measurement scales as a test of whether measurements made by one laboratory are consistent with with those made by another. See [2] for the international protocols. These comparisons involve the measurement of a particular physical characteristic of suitable stable artefacts by the participating laboratories (referred to herein as participants). Measured values are reported together with estimates of variance and covariance. These are analysed to test the hypothesis that each participant has made an appropriate estimate of these quantities.

The variances are estimated by participants based on their knowledge and experience of the measurement system and correspond to two independent error components: a participant-specific component (known to metrologists as the 'systematic' component of error), defined as one that takes the same value for all measurements by a participant, i.e. is fully correlated between all their measurements, and a term encompassing all remaining error (in metrology, referred to as the random component of error). The variance associated with the random component of error is in general derived from a statistical analysis of multiple measurements; the variance associated with the systematic error describes the participant's best estimate of the distribution of possible residual systematic errors (according to the recommendations of the Guide to the Expression of Uncertainty in Measurement [1]) after all known corrections have been applied.

The measurements submitted to a comparison have traditionally been modelled as

$$Y_{\lambda,j,r} = \theta_j + \Delta_\lambda + \varphi_\lambda + \varepsilon_{\lambda,j,r} \tag{1}$$

where $Y_{\lambda,j,r}$ is the $r^{\text{th}}$ measurement made by participant $\lambda$ ($\lambda = 1, \ldots, L$) of artefact $j$ ($j = 1, \ldots, J$); $\theta_j$ is the 'true value' of artefact $j$ which is usually a constant but could include a parametric dependence on other physical quantities such as temperature or time; the random effect $\varphi_\lambda$ is the systematic error of participant $\lambda$ in the measurement due to recognised sources; the random effect $\varepsilon_{\lambda,j,r}$ is the remaining random error in the observation, $Y_{\lambda,j,r}$; and the fixed effect $\Delta_\lambda$ is the participant effect (often referred to as the 'bias' of a participant in metrological literature), representing the component of systematic error in the measurement procedure of participant $\lambda$ that is due to unrecognised sources of error (rather than the total systematic error in the measurement procedure as has sometimes been inferred). Equation (1) is the model implicitly assumed in the usual step-by-step or *ad hoc* analyses of a comparison; it is also the most commonly used model in least squares analyses. The two sets of errors $\{\varphi_\lambda\}$ and $\{\varepsilon_{\lambda,j,r}\}$ are independent of each other, have zero expectation and are drawn from parent distributions whose variances have been estimated by the participant. Typically the covariance structure of the $\{\varepsilon_{\lambda,j,r}\}$ will involve correlations within subsets of a participant's measurements and the covariance structure of the $\{\varphi_\lambda\}$ may involve correlations between two or more participants' measurements. The errors in the estimates of variance and covariance, as supplied to the analyst by the participants, are assumed negligible. The results found in this paper hold whether or not the variances and covariances are known, but if they are subject to non-negligible error, this may need to be taken into account when determining confidence intervals.

Since the 'true values', $\theta_j$, are unknown an indeterminacy exists in the model as a shift in the values of all the $\theta_j$ can be balanced by a shift in the $\Delta_\lambda$ in the opposite direction. To avoid this indeterminacy the participants agree to determine 'consensus' values of the $\theta_j$ by assuming a constraint

$$\sum_{\lambda=1}^{L} w_\lambda \Delta_\lambda = d \qquad (2)$$

on the $\Delta_\lambda$ where $w_\lambda \geq 0$, $\sum_{\lambda=1}^{L} w_\lambda = 1$ and $d$ is a constant. Then, subject to a non-singularity condition, the best linear unbiased estimator (BLUE) of the $\theta_j$ and $\Delta_\lambda$ will be unique and can be found using generalised least squares (GLS). See, for example, [8], sections 3.2 and 3.6. The weights $w_\lambda$ are assigned according to a pre-determined protocol: for example, if one participant is considered to be the 'standard' then its $w_\lambda$ will be one and the rest zero. Alternatively all the participants may be considered equal and so all the $w_\lambda$ will be equal. Or there might be a more complicated weighting. If one or more of the participant effects, $\Delta_\lambda$, is known, then (2) can be used with a non-zero value for $d$ to incorporate this knowledge into the model by way of a weighted average of those known effects. In the absence of prior knowledge of participant effects or artefact values, the constant $d$ is taken to be zero. The particular characteristics of *key comparisons* and the application of GLS to their analysis are discussed in more detail by Koo and Clare [3].

Within the metrological community there have been differences in the way in which the $\varphi_\lambda$ term has been handled due to ambiguity concerning the roles of the $\Delta_\lambda$ and the $\varphi_\lambda$ in the model (1) for the measurement. White [10] and Woolliams *et al.* [11] argue that the unknown effects $\Delta_\lambda$ and the $\varphi_\lambda$ represent the same parameter and hence the $\varphi_\lambda$ should not enter into the estimates of the participant effects although they should contribute to their variances. To accommodate this, White [10] excludes the $\varphi_\lambda$ from the least squares estimator and from the calculation of the covariance matrix associated with the solution but takes them into account by the addition of an extra term to the variance of the estimated participant effects. Woolliams *et al.* [11] likewise exclude the $\varphi_\lambda$ from the covariance matrix in the GLS estimator but do include them in determining the covariance matrix of the estimates. In contrast other authors, e.g. Sutton [9], retain the $\varphi_\lambda$ throughout the GLS calculation. This paper shows that these three approaches give the same results and verifies and generalises White's [10] adjustment for the variances.

It should be noted that estimates of the participant effects, as defined in equation (1), are the quantities required for testing the consistency of participants' reported variance/covariance values. If the measurement results were instead being used to determine the total systematic errors in participants' measurement procedures, then the $\varphi_\lambda$ term should be entirely removed from the model, and the $\Delta_\lambda$ would then represent the total systematic error; the participants' estimates of variance associated with the $\{\varphi_\lambda\}$ do not contribute to the calculated variances of the $\hat{\Delta}_\lambda$.

Rao [7], Zyskind [12], Kruskal [4] and Puntanen & Styan [6] identify conditions under which the GLS estimator is equal to the ordinary least squares (OLS) estimator even though the covariance matrix of the measurements is not a multiple of the identity matrix. In section 2 we consider a model that includes the laboratory comparison, (1), described above. We show that, under a suitable linear transformation, this model meets a condition described by Rao [7], and so the $\varphi_\lambda$ can be ignored when estimating the artefact and participant effect values. We also derive the covariance matrix associated with these estimates. In section 3 we specialise these results for laboratory comparisons, briefly describe a recent key comparison and use it to illustrate the structure of the covariance matrix of the measurements, and verify the expression given by White [10]. Simplified expressions for the GLS estimator and for the covariance matrix associated with the estimates are presented as a corollary in section 4.

**Notation**. We use $X'$ to denote the transpose of a matrix or vector, $X$. The $L_2$ norm of a vector, $\mathbf{v}$, is denoted by $\|\mathbf{v}\|_2$. The inverse of the square root of a positive definite matrix, $X$, obtained using an eigenvalue decomposition is denoted by $X^{-1/2}$. An $i \times j$ matrix of zeros is denoted by $0_{i,j}$ and an $i \times j$ matrix of ones is denoted by $1_{i,j}$. Similarly, $0_i$ and $1_i$ denote column vectors of length $i$ composed of zeros or ones respectively. Where a matrix $X$ is being multiplied by a scalar $c$, we will show the product as $c.X$.

Seber [8] defines a best linear unbiased estimator (BLUE) only for scalars. We define a vector $\mathbf{b}$ as a BLUE for a vector $\boldsymbol{\beta}$ if $\mathbf{g}'\mathbf{b}$ is a BLUE for $\mathbf{g}'\boldsymbol{\beta}$ for every vector $\mathbf{g}$. This is a property of GLS estimators. This definition implies, in particular, that if $\mathbf{b}$ is a BLUE of $\boldsymbol{\beta}$ then $G\mathbf{b}$ is a BLUE of $G\boldsymbol{\beta}$ where $G$ is a matrix.

## 2. Estimation in the presence of a participant-specific random effect

In this section we treat a more general situation than that described in equation (1), both simplifying the presentation and allowing for generalisation. We start with the situation where the constraint (2) is not needed. Consider a linear model with design matrix, $X$, where the observation vector has a covariance matrix, $\Sigma$. Rao [7], Lemma 5a, shows that the ordinary least squares (OLS) estimate gives the same results as the generalised least squares (GLS) solution and hence gives best linear unbiased estimates (BLUE) if and only if $\Sigma = XAX' + ZBZ' + \sigma^2 I$ for some $A$, $B$ and $\sigma^2$ where $X'Z = 0$. In particular, if $\Sigma = XAX' + \sigma^2 I$, OLS will give the same results as GLS.

Suppose the linear model can be written in the form

$$\mathbf{Y} = X\boldsymbol{\beta} + X\boldsymbol{\varphi} + \boldsymbol{\varepsilon} \qquad (3)$$

where $\boldsymbol{\varphi}$ and $\boldsymbol{\varepsilon}$ are vectors of random variables, uncorrelated with each other and having expectation zero and covariance matrices, $A$ and $\sigma^2 I$, respectively. Then the covariance matrix of $\mathbf{Y}$ has the form of $\Sigma$ above and by Rao's lemma [7] the BLUE of $\boldsymbol{\beta}$ is the OLS estimator $\mathbf{b} = (X'X)^{-1}X'\mathbf{Y}$. Also $\text{cov}(\mathbf{b}) = (X'X)^{-1}X'\Sigma X(X'X)^{-1} = A + \sigma^2(X'X)^{-1}$.

This result can be extended to the case where the covariance matrix of $\boldsymbol{\varepsilon}$ is not $\sigma^2 I$ but some positive definite matrix $V_0$. We can transform (3):

$$V_0^{-1/2}\mathbf{Y} = V_0^{-1/2}X\boldsymbol{\beta} + V_0^{-1/2}X\boldsymbol{\varphi} + V_0^{-1/2}\boldsymbol{\varepsilon}. \qquad (4)$$

Then the covariance matrix for $V_0^{-1/2}\mathbf{Y}$ is $\Sigma = V_0^{-1/2} X A X' V_0^{-1/2} + I$ and we have satisfied Rao's condition. See also Rao's equation (68). By Rao's lemma the BLUE is obtained from the OLS estimator for (4) as

$$\mathbf{b} = (X'V_0^{-1}X)^{-1} X'V_0^{-1}\mathbf{Y} \qquad (5)$$

and the covariance matrix of the estimates is

$$\operatorname{cov}(\mathbf{b}) = A + (X'V_0^{-1}X)^{-1}. \qquad (6)$$

Now introduce the constraint (2), initially with $d = 0$.

**Theorem 1**. *Suppose*

$$\mathbf{Y} = X\boldsymbol{\beta} + X\boldsymbol{\varphi} + \boldsymbol{\varepsilon} \qquad (7)$$

*where $\mathbf{Y}$ is an $m \times 1$ vector of observable random variables, $X$ is an $m \times n$ design matrix, $\boldsymbol{\beta}$ is an $n \times 1$ vector of the unknown parameters, $m$ is the total number of measurements, $\boldsymbol{\varphi}$ is an $n \times 1$ vector of random variables with covariance matrix $A$ and $\boldsymbol{\varepsilon}$ is an $m \times 1$ vector of random variables with non-singular covariance matrix $V_0$. Suppose the $\boldsymbol{\beta}$ are subject to a constraint*

$$\mathbf{w}'\boldsymbol{\beta} = 0 \qquad (8)$$

*where $\mathbf{w}$ is a non-zero $n \times 1$ vector. Suppose $X$ has rank $n-1$ and*

$$\begin{pmatrix} X \\ \mathbf{w}' \end{pmatrix} \qquad (9)$$

*has rank $n$. Then the best linear unbiased estimator of $\boldsymbol{\beta}$ is the same as that for the model (7) with the random effects term, $X\boldsymbol{\varphi}$, omitted.*

**Proof.** We need to transform the model to a form where Rao's [7] lemma can be applied. Let $\mathbf{v} = \mathbf{w}/\|\mathbf{w}\|_2$ and introduce an $n \times (n-1)$ matrix $S$ such that $(\mathbf{v} \ \ S)$ is an orthogonal matrix. It follows that

$$\mathbf{v}\mathbf{v}' + SS' = I, \ \mathbf{v}'\mathbf{v} = 1, \ S'S = I, \ \mathbf{v}'S = 0 \text{ and } S'\mathbf{v} = 0. \qquad (10)$$

Also

$$\begin{pmatrix} I & -X\mathbf{v} \\ 0 & 1 \end{pmatrix} \begin{pmatrix} X \\ \mathbf{v}' \end{pmatrix} (S \ \ \mathbf{v}) = \begin{pmatrix} XS & 0 \\ 0 & 1 \end{pmatrix}$$

has rank $n$, since the first and last matrices in the left-hand expression are non-singular, and hence $XS$ has rank $n-1$. Any $\boldsymbol{\beta}$ satisfying (8) can be written $\boldsymbol{\beta} = S\breve{\boldsymbol{\beta}}$ where $\breve{\boldsymbol{\beta}} = S'\boldsymbol{\beta}$ is an $(n-1) \times 1$ vector since $SS'\boldsymbol{\beta} = (I - \mathbf{v}\mathbf{v}')\boldsymbol{\beta} = \boldsymbol{\beta}$. And any such $\boldsymbol{\beta}$ satisfies the constraint. So there is a one-to-one correspondence between the values of $\breve{\boldsymbol{\beta}}$ and those values of $\boldsymbol{\beta}$ that satisfy the constraint. Thus, the model (7) can be rewritten as

$$\mathbf{Y} = XS\breve{\boldsymbol{\beta}} + X\boldsymbol{\varphi} + \boldsymbol{\varepsilon}. \qquad (11)$$

Since $XS$ has rank $n-1$, equalling the number of columns, the system is uniquely solvable for $\breve{\boldsymbol{\beta}}$ and hence for $\boldsymbol{\beta}$.

We now show that the $X\boldsymbol{\varphi}$ term is of the form $XS\breve{\boldsymbol{\varphi}}$ so that (11) is of the form (3). Since $X$ has rank $n-1$ there exists a non-zero $m \times 1$ vector $\mathbf{f}$ such that

$$X\mathbf{f} = 0. \tag{12}$$

Also

$$\begin{pmatrix} X \\ \mathbf{v}' \end{pmatrix} \mathbf{f} \neq 0$$

so that $\mathbf{v}'\mathbf{f} \neq 0$. Let

$$F = (I - \mathbf{f}\mathbf{v}'/\mathbf{v}'\mathbf{f}) = (I - \mathbf{f}\mathbf{w}'/\mathbf{w}'\mathbf{f}) \tag{13}$$

and $T = S'F$. Then $ST = SS'F = (I - \mathbf{v}\mathbf{v}')F = F$ so that $XST = XF = X$ and hence

$$\mathbf{Y} = XS\breve{\boldsymbol{\beta}} + XS\breve{\boldsymbol{\varphi}} + \boldsymbol{\varepsilon}, \tag{14}$$

where $\breve{\boldsymbol{\varphi}} = T\boldsymbol{\varphi}$, which is in the form of (3). Hence the $XS\breve{\boldsymbol{\varphi}}$ can be omitted from (14) when calculating the best linear unbiased estimator of $\breve{\boldsymbol{\beta}}$ and so $X\boldsymbol{\varphi}$ can be omitted from (7). This completes the proof. □

From (5) and (6) the best linear unbiased estimator of $\breve{\boldsymbol{\beta}}$ is $\breve{\mathbf{b}} = (S'X'V_0^{-1}XS)^{-1}S'X'V_0^{-1}\mathbf{Y}$ and $\text{cov}(\breve{\mathbf{b}}) = TAT' + (S'X'V_0^{-1}XS)^{-1}$. Hence the best linear unbiased estimator of $\boldsymbol{\beta}$ is

$$\mathbf{b} = S(S'X'V_0^{-1}XS)^{-1}S'X'V_0^{-1}\mathbf{Y} \tag{15}$$

and its covariance matrix is

$$\text{cov}(\mathbf{b}) = STAT'S' + S(S'X'V_0^{-1}XS)^{-1}S' = FAF' + S(S'X'V_0^{-1}XS)^{-1}S'. \tag{16}$$

The second term, for which a simplification is derived in section 4, is just the covariance matrix when there is no $X\boldsymbol{\varphi}$ term and the first term is the adjustment when the $X\boldsymbol{\varphi}$ term is present. This gives the following corollary.

**Corollary 1.1.** *Under the conditions of Theorem 1 the covariance matrix of the best linear unbiased estimator must be increased by $FAF'$, where $F$ is defined in (13) when the random effects term $X\boldsymbol{\varphi}$ is included in the model.*

Now consider the case where we have the constraint $\mathbf{w}'\boldsymbol{\beta} = d$ with non-zero $d$ as in (2). We can rewrite (7) as $\mathbf{Y} - \breve{d}.X\mathbf{w} = X(\boldsymbol{\beta} - \breve{d}.\mathbf{w}) + X\boldsymbol{\varphi} + \boldsymbol{\varepsilon}$ where $\breve{d} = d/\|\mathbf{w}\|_2^2$. Now $\mathbf{w}'(\boldsymbol{\beta} - \breve{d}.\mathbf{w}) = 0$ so if we replace $\mathbf{Y}$ by $\mathbf{Y} - \breve{d}.X\mathbf{w}$ and $\boldsymbol{\beta}$ by $\boldsymbol{\beta} - \breve{d}.\mathbf{w}$ we have transformed the problem to that considered in Theorem 1. So we have the following corollary.

**Corollary 1.2.** *If (8) in Theorem 1 is replaced by $\mathbf{w}'\boldsymbol{\beta} = d$ the results of the theorem and Corollary 1.1 are unchanged.*

## 3. Measurement Comparisons

We apply the results of the previous section to a measurement comparison, (1). An example of such a comparison exhibiting considerable complexity is shown in Figure 1. The diagram shows the exchange of 7 artefacts between 15 laboratories. The numbers beside arrowheads are numbers of measurements of an artefact by a laboratory in each round of artefact exchange. The covariance structure of the measurements can be demonstrated by considering possible correlations between measurements indicated in the diagram. Suppose that some component of MSL's measurement system had been calibrated at NIST. Then the error in all of MSL's measurements due to that component would be correlated with those of NIST, i.e. the error $\varphi_{MSL}$ would be correlated with the error $\varphi_{NIST}$. The matrix $A$ of (6) would then have a corresponding non-zero off-diagonal element. Similarly, one might imagine that of the four measurements made by NIST on artefact E, the three done before the artefact was sent to MSL were subject to a common error which was not shared by the final measurement made after the artefact was returned. In this case the $V_0$ matrix would include non-zero off diagonal elements within the block corresponding to the measurements made by NIST describing the correlation between the errors for the first three measurements.

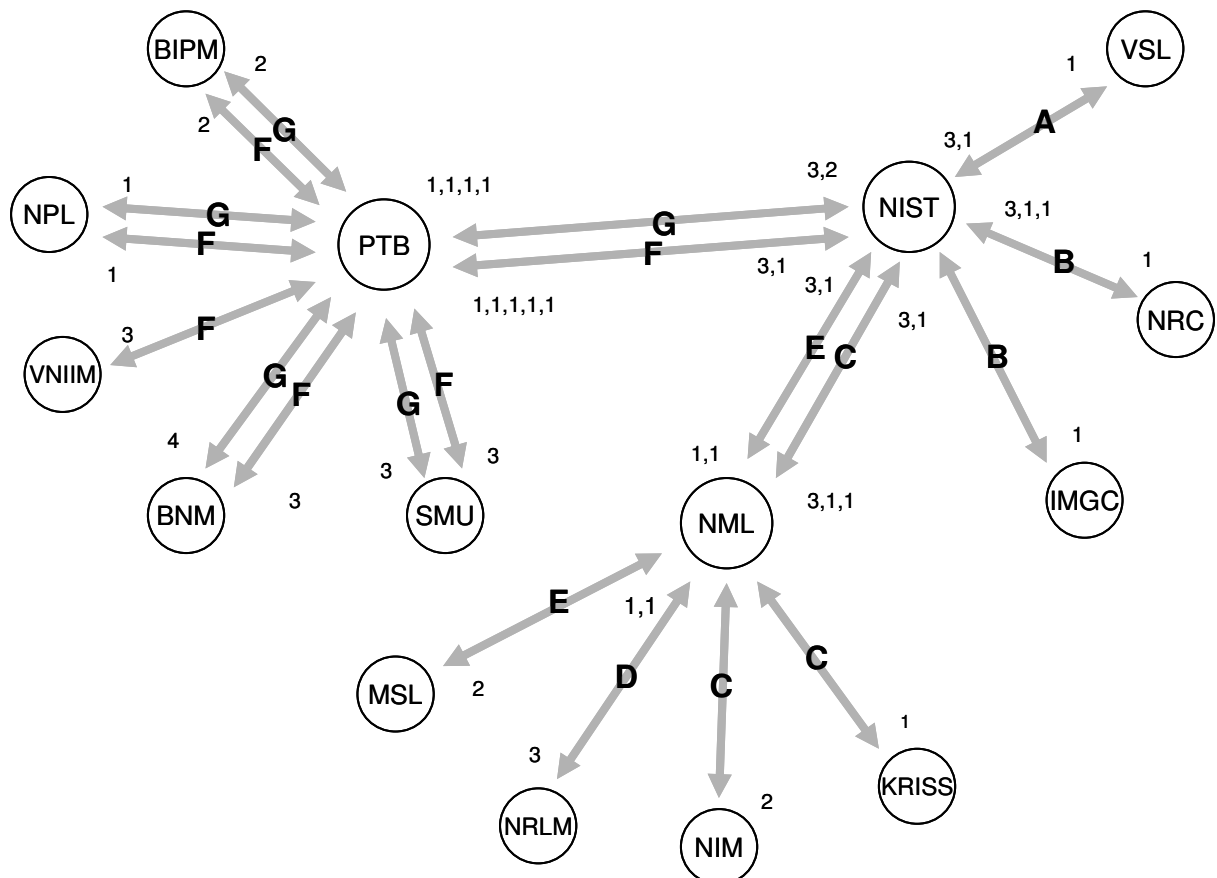

Figure 1. Diagram showing the exchange of artefacts A…G in a comparison involving a coordinator (NIST), two sub-coordinators (NML and PTB) and 12 other laboratories. In this case the artefacts were standard platinum resistance thermometers which the participating laboratories were required to use to measure their realization of the gallium fixed point on the ITS-90 temperature scale [5].

Suppose that the design matrix, $X$, is organised so that the first $J$ elements of $\boldsymbol{\beta}$ denoted by $\boldsymbol{\Theta}$ represent the artefact values and the remaining $L$ elements denoted by $\boldsymbol{\Delta}$ denote the participant values so that $\boldsymbol{\beta}' = (\boldsymbol{\Theta}' \quad \boldsymbol{\Delta}')$. The elements of each of the rows of $X$ are zero apart from two elements equal to unity, of which the first is in one of the first $J$ columns and the other is in one of the remaining $L$ columns. These assign the appropriate values of $\boldsymbol{\Theta}$ and $\boldsymbol{\Delta}$ to each of the measurements in the $\mathbf{Y}$ vector. In this way, (1) can be represented by (7). Provided the full rank condition (9) holds, Theorem 1 can be applied and so we can ignore the $\varphi_\lambda$ term when estimating the values of $\boldsymbol{\beta}$ as White [10] and Woolliams *et al.* [11] propose. Koo & Clare [3] show that a necessary and sufficient condition for (9) to hold is that any two artefacts in the comparison are linked through a sequence of measurements of overlapping pairs of artefacts.

Now look at the adjustment to the covariance matrix, $FAF'$. The random effects apply only to the participant effects, i.e. the first $J$ elements of $\boldsymbol{\varphi}$ are zero, so we can write

$$A = \begin{pmatrix} 0_{J,J} & 0_{J,L} \\ 0_{L,J} & \breve{A} \end{pmatrix}$$

where $\breve{A}$ is the covariance matrix of the random effects, $\varphi_\lambda$. The constraint (8) applies only to the participants so we can put $\mathbf{w}' = (0'_J \quad \breve{\mathbf{w}}')$. We can choose $\mathbf{f}' = (-1'_J \quad 1'_L)$. Since $\mathbf{w}'\mathbf{f} = 1$ we have

$$F = \begin{pmatrix} I & 1_J \breve{\mathbf{w}}' \\ 0_{L,J} & I - 1_L \breve{\mathbf{w}}' \end{pmatrix}.$$

Multiplying out $FAF'$, we obtain the following theorem.

**Theorem 2.** *In the laboratory comparison situation described above the adjustment to the covariance term can be evaluated as follows:*

$$FAF' = \begin{pmatrix} B & C \\ C' & D \end{pmatrix}$$

*where* $B = 1_J \breve{\mathbf{w}}' \breve{A} \breve{\mathbf{w}} 1'_J = \breve{\mathbf{w}}' \breve{A} \breve{\mathbf{w}}.1_{JJ}$, $C = 1_J \breve{\mathbf{w}}' \breve{A} (I - \breve{\mathbf{w}} 1'_L)$ *and* $D = (I - 1_L \breve{\mathbf{w}}') \breve{A} (I - \breve{\mathbf{w}} 1'_L)$.

The last term, $D$, is the adjustment to the covariance term of the estimates of the participant effects and so is the term of most interest. Expressing this in terms of elements:

$$D_{\lambda,\mu} = \breve{A}_{\lambda,\mu} - \sum_{\xi=1}^{L} \breve{A}_{\lambda,\xi} \breve{w}_\xi - \sum_{\gamma=1}^{L} \breve{w}_\gamma \breve{A}_{\gamma,\mu} + \sum_{\gamma=1}^{L} \sum_{\xi=1}^{L} \breve{w}_\gamma \breve{A}_{\gamma,\xi} \breve{w}_\xi .$$

In particular, if $\breve{A}$ is diagonal, that is the random effects, $\varphi_\lambda$, are uncorrelated with each other, then the diagonal elements of $D$ are

$$D_{\lambda,\lambda} = \breve{A}_{\lambda,\lambda} - 2\breve{w}_\gamma \breve{A}_{\lambda,\lambda} + \sum_{\gamma=1}^{L} \breve{w}_\gamma^2 \breve{A}_{\gamma,\gamma} = (1 - \breve{w}_\gamma)^2 \breve{A}_{\lambda,\lambda} + \sum_{\substack{\gamma=1 \\ \gamma \neq \lambda}}^{L} \breve{w}_\gamma^2 \breve{A}_{\gamma,\gamma}.$$

This is the expression obtained by White ([10], equation 16) for a comparison in which one artefact is measured once by each participant and in which their errors are uncorrelated.

There may be other predictors or covariates. For example, if measurements are made at different temperatures (1) becomes

$$y_{\lambda,j,r} = \theta_j + \kappa(T_{\lambda,j,r} - T_0) + \Delta_\lambda + \varphi_\lambda + \varepsilon_{\lambda,j,r}$$

where $T_{\lambda,j,r} - T_0$ is the deviation of the temperature from the reference value and the unknown $\kappa$ is an additional element in $\boldsymbol{\beta}$. To allow for this and similar situations, suppose that there are additional elements in $\boldsymbol{\beta}$ with corresponding zero elements in $\mathbf{w}$, $\mathbf{f}$ and $A$. Then the results of this section are unchanged and there is no correction to the variances for these additional predictors provided that the rank of $X$ remains at $n-1$, i.e. one less than the number of columns.

## 4. Simplification of b and cov(b)

The GLS solution can be found by reducing the problem to a non-singular one by using the constraint (2) to eliminate one of the $\Delta_\lambda$ (see, for example, Woolliams *et al.* [11]). This section presents two alternative methods. Let $F$, $S$, $T$, $\mathbf{f}$ and $\mathbf{v}$ be as in the proof of Theorem 1. Then $TS = I$ and so

$$\begin{pmatrix} T \\ \mathbf{v}' \end{pmatrix} (S \quad \mathbf{v}) = \begin{pmatrix} I & T\mathbf{v} \\ 0 & 1 \end{pmatrix}$$

is non-singular and therefore $(T' \quad \mathbf{v})'$ is also non-singular. The component of the covariance matrix of $\mathbf{b}$ arising from $\boldsymbol{\varepsilon}$ in (16) can be written:

$$S(S'X'V_0^{-1}XS)^{-1}S' = (S \quad 0) \begin{pmatrix} S'X'V_0^{-1}XS & 0 \\ 0 & c \end{pmatrix}^{-1} \begin{pmatrix} S' \\ 0 \end{pmatrix} \quad \text{for any } c \neq 0$$

$$= (S \quad 0) \begin{pmatrix} T \\ \mathbf{v}' \end{pmatrix} \left\{ (T' \quad \mathbf{v}) \begin{pmatrix} S'X'V_0^{-1}XS & 0 \\ 0 & c \end{pmatrix} \begin{pmatrix} T \\ \mathbf{v}' \end{pmatrix} \right\}^{-1} (T' \quad \mathbf{v}) \begin{pmatrix} S' \\ 0 \end{pmatrix} \quad (17)$$

$$= ST(T'S'X'V_0^{-1}XST + c.\mathbf{v}\mathbf{v}')^{-1}T'S'$$

$$= STP^{-1}T'S'$$

where $P = X'V_0^{-1}X + c.\mathbf{v}\mathbf{v}'$. The estimator (15) can be written

$$\mathbf{b} = S(S'X'V_0^{-1}XS)^{-1}S'X'V_0^{-1}\mathbf{Y} = STP^{-1}T'S'X'V_0^{-1}\mathbf{Y} = STP^{-1}X'V_0^{-1}\mathbf{Y}. \quad (18)$$

Since $X\mathbf{f} = \mathbf{0}$ we have $P\mathbf{f} = c.\mathbf{v}\mathbf{v}'\mathbf{f} = c\mathbf{v}'\mathbf{f}.\mathbf{v}$. From (17), $P$ is non-singular and so $P^{-1}\mathbf{v} = \mathbf{f}/(c\mathbf{v}'\mathbf{f})$ and $P^{-1}\mathbf{v}\mathbf{f}' = \mathbf{f}\mathbf{f}'/(c\mathbf{v}'\mathbf{f})$. Hence $P^{-1}F' = P^{-1} - \mathbf{f}\mathbf{f}'/\{c(\mathbf{v}'\mathbf{f})^2\} = FP^{-1}$ by symmetry and so $FP^{-1}F' = P^{-1}F'^2 = P^{-1}F'$ since $F^2 = F$. Thus, we can further simplify (17) and (18):

$$S(S'X'V_0^{-1}XS)^{-1}S' = FP^{-1}F' = P^{-1}F' = FP^{-1} = P^{-1} - \mathbf{f}\mathbf{f}'/\{c(\mathbf{v}'\mathbf{f})^2\}, \quad (19)$$

$$\mathbf{b} = FP^{-1}X'V_0^{-1}\mathbf{Y} = P^{-1}F'X'V_0^{-1}\mathbf{Y} = P^{-1}X'V_0^{-1}\mathbf{Y}. \quad (20)$$

These expressions will hold for any $c \neq 0$. One can confirm that $P^{-1}F'$ is, indeed, not dependent on the value of $c$ by adding an extra term $g.\mathbf{v}\mathbf{v}'$ to $P$ and expanding the inverse using the Sherman-Morrison formula. It is convenient to rescale $c$ so that $P = X'V_0^{-1}X + c.\mathbf{ww}'$. Then we have the following corollary.

**Corollary 1.3.** *Given the model and conditions of Theorem 1 and $c \neq 0$ we can obtain the best linear unbiased estimator (BLUE) of the unknowns unambiguously from (20) as*

$$\mathbf{b} = (X'V_0^{-1}X + c.\mathbf{ww}')^{-1} X'V_0^{-1}\mathbf{Y} \qquad (21)$$

*and the covariance matrix of that estimate from (16) and (19) as*

$$\begin{aligned} \text{cov}(\mathbf{b}) &= FAF' + (X'V_0^{-1}X + c.\mathbf{ww}')^{-1} F' \\ &= FAF' + (X'V_0^{-1}X + c.\mathbf{ww}')^{-1} - \mathbf{ff}'/\{c(\mathbf{w'f})^2\} \end{aligned} \qquad (22)$$

*where $\mathbf{f}$ and $F$ are as defined in (12) and (13).*

We can interpret the term $X'V_0^{-1}X + c.\mathbf{ww}'$ that occurs in (21) and (22) as follows. Add another 'observation' to the data by appending a row $c_1.\mathbf{w}'$ to $X$, and appending a zero, or $c_1 d$ if we want a bias as in equation (2), to $\mathbf{Y}$. Add an extra row and column to $V_0$ with $c_2$ in the bottom right hand corner and zeros elsewhere. Then carry out a generalised least squares analysis as in equation (5). This gives (21) with $c = c_1^2/c_2$. Similarly the resulting covariance matrix needs to be adjusted as in (22). The scalars $c_1$ and $c_2$ are arbitrary apart from being non-zero but for numerical stability should be chosen to be of a similar size to the elements of $X$ and $V_0$ respectively. This method has the advantage of not requiring a change in the number of columns of $X$ during the calculation.

There is another approach to calculating (21) and (22). Suppose we have already found an estimate $\mathbf{b}_0$ of $\boldsymbol{\beta}$ with a constraint vector $\mathbf{w}_0$. Then we can use it to determine an estimator $\mathbf{b} = (I - \mathbf{fw}'/\mathbf{w'f})\mathbf{b}_0 = F\mathbf{b}_0$ that satisfies the constraint $\mathbf{w'b} = 0$. It follows from (12) that the addition of the extra term $\mathbf{fw'b}_0/\mathbf{w'f} = (\mathbf{w'b}_0/\mathbf{w'f}).\mathbf{f}$ will not change the goodness of the fit. If we need the constraint as in (2) add an extra term $(d/\mathbf{w'f}).\mathbf{f}$ to $\mathbf{b}$. The covariance matrix for $\mathbf{b}$ is obtained by pre- and post-multiplying the covariance matrix for $\mathbf{b}_0$ by $F$ and $F'$ respectively. This is a possibly convenient computational method since one can initially constrain one of the elements of $\mathbf{b}_0$ to be zero and then apply the correction.

Another version of the formulae (21) and (22) can be found by replacing $V_0$ by the full covariance matrix $V_0 + XAX'$ and setting $A$ to zero in (22).

## 5. Discussion

Three GLS formulations found in the literature [9, 10, 11] for use in the analysis of measurement comparisons have been shown here to be equivalent. The metrological community can therefore use any of these implementations without ambiguity with regard to the results for the purpose of testing the consistency of a participant's measurements with the reported variances and covariances. An alternative equivalent formulation, requiring no transformation of the design matrix to incorporate the constraint, is also given here.

While the result shown here is particularly important in the context of comparison analysis, the equivalence between the full GLS operator and the GLS operator which excludes components of variance correlated between all measurements of any participant will also hold in any other context with the same structure as the comparison problem. In any such context, this result shows the insensitivity of the values of unknowns to those correlated components of variance and also allows the separation of their contribution to the variances of the unknowns.

**Acknowledgements**

The authors thank D R White, R Willink, D Krouse and C M Sutton for helpful discussions and D. R. White for Figure 1. This work was funded by the New Zealand Government under a contract for the provision of national measurement standards.